\documentclass[aps,a4paper,10pt,twocolumn,nofootinbib,eqsecnum]{revtex4} 
\usepackage[T1]{fontenc}
\usepackage{mathptmx}
\usepackage{datetime}
\usepackage{amsmath}
\usepackage{amsfonts}
\usepackage{amssymb}
\usepackage{mathrsfs}
\usepackage[mathscr]{euscript}
\usepackage{mathtools}
\usepackage{textcomp} 
\usepackage[dvips]{graphicx}
\usepackage{fancyhdr}
\usepackage{colordvi}
\usepackage{epsfig}
\usepackage{color}
\usepackage{bm}

\def\overstrike#1#2{{\setbox0\hbox{$#2$}\hbox to \wd0{\hss
    $#1$\hss}\kern-\wd0\box0}}

\renewcommand{\Vec}{\textbf}

\newdateformat{yymmdddate}{\THEYEAR/\twodigit{\THEMONTH}/\twodigit{\THEDAY}}

\begin{document}
\title{Time and space, frequency and wavevector: \\
\normalsize{or, what I talk about when I talk about propagation}}
\author{Paul Kinsler}
\email{Dr.Paul.Kinsler@physics.org}
\affiliation{
  Blackett Laboratory, Imperial College London,
  Prince Consort Road,
  London SW7 2AZ, 
  United Kingdom.
}

\lhead{\includegraphics[height=5mm,angle=0]{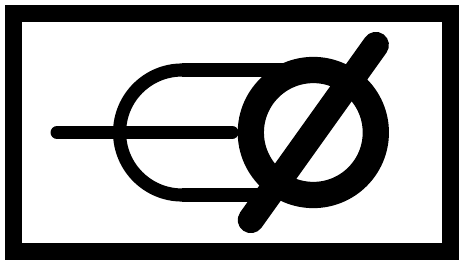}~~NEGFRQ}
\chead{Time and space, frequency and wavevector}
\rhead{
\href{mailto:Dr.Paul.Kinsler@physics.org}{Dr.Paul.Kinsler@physics.org}\\
\href{http://www.kinsler.org/physics/}{http://www.kinsler.org/physics/}
}

\begin{abstract}

The existence of waves with negative frequency
 is a surprising and perhaps controversial claim 
 which has recently been revisted in optics
 and for water waves.
Here I explain a context within which to understand the 
 meaning of the ``negative frequency'' conception, 
 and why it appears in some cases.

\end{abstract}

\date{\today}
\maketitle
\thispagestyle{fancy}

%

%
\section*{Introduction}\label{S-intro}

Claims or analyses involving negative frequency waves
 occur in a wide range of fields:
 quantum mechanics, 
 optics \& EM \cite{Rubino-MKBTRKLKF-2012prl,Conforti-WBTF-2013pra},
 and acoustics and water waves \cite{Rousseaux-MMPL-2008njp}.
However, 
 some regard the notion of negative frequencies as unphysical.
Here I describe 
 four alternative views of the universe
 with a view to clarifying why negative frequencies
 might appear, 
 and whether or not we should be worried:
  the \emph{omniscient}, 
  the \emph{traditional}, 
  the \emph{directed}, 
  and the \emph{causal}.

The ``traditional'' and the ``directed'' pictures, 
 involving propagation along time axis
 and one selected spatial axis respectively, 
 have already been compared
 for a range of acoustic wave equations \cite{Kinsler-2012arXiv-fbacou}.
In particular, 
 the convenience of spatial propagation in optics, 
 and the additional utility of the unidirectional approximation
 has been widely studied 
 (see e.g. \cite{Kinsler-2010pra-fchhg} and references therein).
Here I am most interested in 
 comparing the traditional \& directed views, 
 and restrict notation to Cartesian coordinates 
 for simplicity; 
 although extension to alternate systems is of course possible
 \cite{Kinsler-2012arxiv-fbrad}.
It is important to note that in practical terms,
 utilising spatial propagation amounts to making an approximation, 
 since not only the initial conditions 
 but the ongoing propagation 
 require a knowledge of the future which we rarely have.

It is useful to mention two ideas before we start:
 first, 
 the standard notion of the \emph{casual past} of a point in time and space, 
 generally called the past light cone; 
 and second, 
 the \emph{computational past} of that point.
This notion of computational past
 applies primarily to numerical simulations, 
 but can also be useful from more abstract perspective.
It consists of
 both the set of initial conditions 
 and/or computed data
 which might have affected the calculated state of that chosen point.

In what follows I will denote the profile of the wave field under discussion
 as $F$, 
 but rather than festooning it with a plethora of typographical tics
 (bars, tildes, primes, and the like)
 to indicate what arguments may or may not have been transformed,
 I use its arguments alone to define $F$'s meaning --
 whether time $t$ or space $\Vec{r}=(x,y,z)$, 
 whether it is a spectra with frequency $\omega$
 or wavevector $\Vec{k}=(k_x,k_y,k_z)$, 
 and so on.
Further, 
 I will not consider any specifics as to why a specified wave field
 might develop a certain character or characteristics -- 
 pulses, oscillations, spectral features and the like -- 
 just what we can say about the wave
 given the available knowledge.

In addition to the four sections addressing each of the 
 omniscient, traditional, directed, and causal pictures in turn, 
 along with a brief conclusion, 
 there are appendices which address some issues 
 relevant in particular to the field of nonlinear optics:
 the handling of time-response models, 
 and the appearance of negative frequency terms.



%
\section{Omniscient: All time and all space}\label{S-timespace}

The \emph{omniscient} position is taken when we claim
  all possible knowledge about how some wave has and will move
  through its environment.
Here we might express our knowledge of some wave-field $F$
  by writing it with full space and time arguments: i.e. $F(t,\Vec{r})$.
Most likely, 
 this view is a result of us having an analytic solution 
 to some specific case of the physical model we are interested in.

Given this fully known wave field solution $F(t,\Vec{r})$, 
 we are at liberty to Fourier transform in either time or space
 (or both) 
 as we see fit.
In the resulting double spectrum ${F}(\omega,\Vec{k})$,
 we will quite naturally see
 not only positive and negative frequency components,
 but also forward and backward wavevector components
 along each spatial coordinate axes.

Since the spectrum of some function is closely related
 to the complex conjugate of that spectrum for negative frequencies, 
 we know for the doubly transformed 
 $F(\omega,\Vec{k})$ 
 that $F(\omega,\Vec{k}) = F(-\omega,-\Vec{k})$.
Notably, 
 in this case 
 there is no obvious distinction between
 a positive frequency disturbance evolving in one direction
 and
 a negative frequency disturbance evolving in the opposite direction.
We therefore need not be troubled by the presence of negative frequencies, 
 since we can reinterpret them as oppositely directed positive ones --
 we might, 
 for example, 
 refer to the idea that a positron
 is only an electron travelling backwards in time\footnote{Of course
  the handling of most particles is complicated by their mass, 
  whereas photons, 
  being massless, 
  can be their own antiparticle.}.

It is also possible to consider a ``limited omniscience'' position, 
 where we claim complete knowledge of the wave behaviour in some
 defined region of space and time.
This might be from some (past) experimental results, 
 or be a numerical or analytic solution valid over that limited extent.
In such a case, 
 both time and space can be Fourier transformed, 
 and the remarks above about handling and/or interpreting
 negative frequencies still hold.
However, 
 whilst such after-the-fact analysis is typically very useful, 
 caution should be taken when applying the conclusions
 to an on-going propagation:
 in such inescapably dynamical situations,  
 the alternatives discussed in the next three sections
 are more applicable.


%
\section{Traditional: temporal propagation}\label{S-time}

\begin{figure}
\includegraphics[angle=-0,width=0.82\columnwidth]{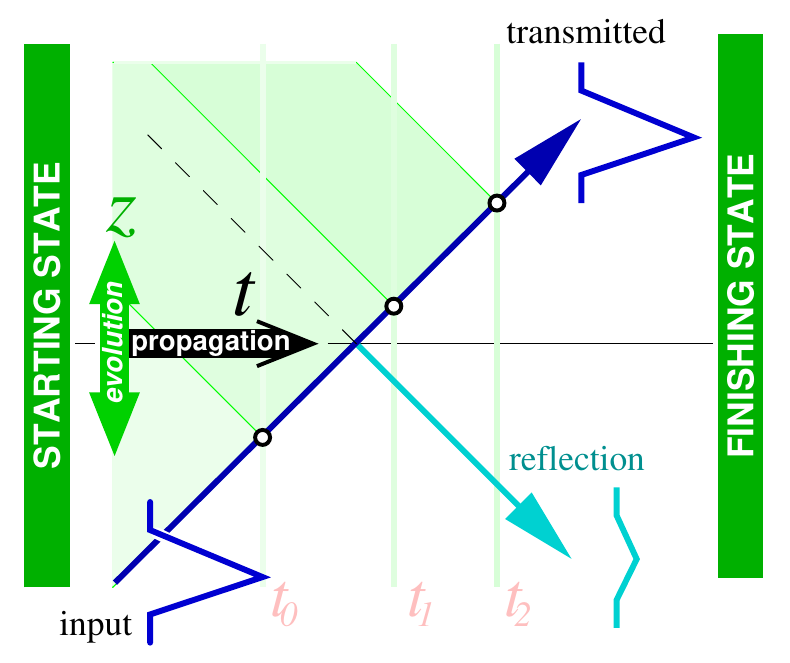}
\caption{
 Temporal propagation of waves, 
  where disturbances (or pulses)
  evolve either forward or backward in space.
 At any point in the propagation, 
  we know the spatial behaviour of our wave field at all points in space, 
  as indicated by the pale green vertical lines.
 The pale green shaded triangles indicate the past light cones
  (i.e. the causal past)
  of a wave element (black circles)
  at selected times  
  along the path of the disturbance.
 In a temporally propagated numerical simulation
  which has a maximum wave speed,
  the causal past matches the computational past.
 A notional interface has been added to the diagram
  to show how a reflection would behave.
}
\label{fig-reflect-t}
\end{figure}

The \emph{traditional} position is taken when we claim
 all knowledge about how the wave has moved 
 through its environment up until now.
This picture is shown on fig. \ref{fig-reflect-t}, 
 and assumes the normal preferred direction of propagation along 
 the time axis, 
 which is usually from $t=-\infty$, 
 through $t=0$, 
 and towards $t=+\infty$.

Here we might express our knowledge of the wave-field $F$
 by writing it at the current time $t_0$ with full space arguments: 
 i.e. $F_{t_0}(\Vec{r})$; 
 and where is is implied that we also knew $F_t(\Vec{r})$
 at all past times $t<t_0$ also.
The starting state of our propagation matches 
 the early-time boundary conditions -- 
 i.e. what would normally be considered to be the initial conditions.
An electromagnetic FDTD simulation \cite{Yee-1966tap,Oskooi-RIBJJ-2010cpc}, 
 along with many other finite element approaches, 
 uses this traditional approach; 
 it is also possible to apply factorization methods 
 to get unidirectional versions of temporally propagated 
 wave equations \cite{Kinsler-2018jo-d2owe}.
In this sense the starting state for some temporally propagated simulation --
 the ``computational initial conditions'' --
 are the same as the traditional physical initial conditions\footnote{However, 
  if we were to decide to propagate backwards in time
  from $t=+\infty$ to $t=-\infty$, 
  as is sometimes done when we have preferred final-time boundary conditions, 
  note that those 
  become the computational initial conditions, 
  and that the computational past of any point in the propagation 
  will be its \emph{future} light cone.
 The interested reader might try drawing a suitable counterpart
  to fig. \ref{fig-reflect-t} themselves.}.

Given some wave field state $F_t(\Vec{r})$ known
 over all space $\Vec{r}$
 at some specified time $t=t_0$,
 we are at liberty to Fourier transform in space, 
 but not time.
The spatial transform will give us the wavevector 
 or momentum-like properties of the wave field,
 in a spectrum ${F_{t_0}}(\Vec{k})$.
This will contain both forward and backward wavevector components
 along each spatial coordinate axis, 
 and we can calculate it not only at the current time $t_0$, 
 but also all past times $t<t_0$.

For simplicity, 
 imagine we are following some trajectory through time and space, 
 while counting interesting events that occur locally.
Time is always increasing but we might travel any direction through space, 
 although it is easier to think of the case where our position is fixed.
Naturally, 
 our count of interesting events will only ever increase, 
 thus any estimate of the time period $T$ between events
 would be a positive number; 
 as would any frequency estimation $f$ in events-per-second.
However, 
 the events themselves may have different spatial characteristics:
 notably, 
 we may see objects that pass us travelling left, 
 or perhaps travelling right.
The differing character of events, 
 accessible to us through our spatial knowledge $F_t(\Vec{r})$,
 allows us to impose a sign (or signs)
 when adjusting our counting total -- 
 perhaps +1 for left-going objects, 
 and -1 for right-going ones.

Having discussed how a counting/timing argument
 that gives us a strictly positive frequency estimate
 can be converted into a signed count using spatial information, 
 let us revist the temporally propagated spatial spectrum ${F_{t_0}}(\Vec{k})$
 of our notional wave field.
As already noted, 
 the spatial spectrum of the wave has both
 positive and negative wavevector components.
Starting with the well known velocity relation for waves $v=\omega/k$, 
 we can use this to claim that the positive wavevector part
 of the spectrum represents 
 forward evolving waves (with $v > 0$), 
 while the negative wavevector part represents 
 backward evolving waves (with $v < 0$).
However, 
 we have to be careful:
 even a static field profile will have positive 
 and negative wavevector components, 
 since for real-valued $F_t(z)$, 
 $F_t(-k)=F_t^\ast(k)$.
It is the changing complex phase(s) of $F_t(k)$ that represents 
 the shift in $F_t(z)$ profile from one position to another.

Note that I have so far only discussed frequency \emph{estimates}.
Strictly speaking, 
 in this picture, 
 any system state exists only at an instant in time, 
 so of itself that state has no frequency content at all.
Further, 
 even given our knowledge of past states,
 we \emph{cannot} calculate a true frequency dependence, 
 because we do not have the entire wave history to hand:
 we have the past but not the future.
The best we might do in getting an up-to-date idea of the spectrum 
 is apply a Laplace transform\footnote{Although any one-sided transform 
  that seemed appropriate, 
  if applied over past (known) data,
  would also be fine.}
 over the known past information, 
 converting $t$ into a Laplace-spectral $s$, 
 and get a hybrid spectrum ${F'_{t_0}}(s,\Vec{k})$.
Nevertheless, 
 it may be possible to characterise the dominant frequency-like 
 properties of the propagation as a function of the know wavevector; 
 and so be able to use a reference frequency $\Omega(\Vec{k})$
 \cite{Kinsler-2012arXiv-fbacou}
 as a basis on which to simplify the propagation -- 
 perhaps by assuming it to be unidirectional.



However, 
 if we judge that omitting the future behaviour
 will have a negligible effect, 
 or that it is irrelevant since we only care about 
 an experiment  or computation that has already finished, 
 we can simply take data from the appropriate past time interval, 
 and calculate the frequency spectrum of that.
This after-the-fact analysis of past data
 returns us to a version of the omniscient position,
 albeit one limited in scope, 
 so that the same view of negative frequencies as there can be applied.
This is what is usually done
 (either implicitly or explicitly), 
 which is of course the reason why so many frequency responses \& spectra 
 appear in the literature, 
 textbooks, 
 and on datasheets.

So in this \emph{traditional} picture, 
 we need not be overly troubled by negative frequencies.
Either we are being very rigorous, 
 and so deny that a true frequency spectrum exists at all, 
 or we have a spectrum calculated from known historical data, 
 which has negative frequencies reinterpretable as positive ones, 
 just as in the omniscient picture.
As a final note, 
 and as a result of the considerations above, 
 the spectroscopists habit of giving spatial 
 (wavevector, or momentum-like) spectra 
 rather than temporal (frequency, or energy-like) ones 
 makes sense from a theoretical perspective,
 as well as from a practical experimental one.

%
\section{Directed: spatial propagation}\label{S-space}

\begin{figure}
\includegraphics[angle=-0,width=0.82\columnwidth]{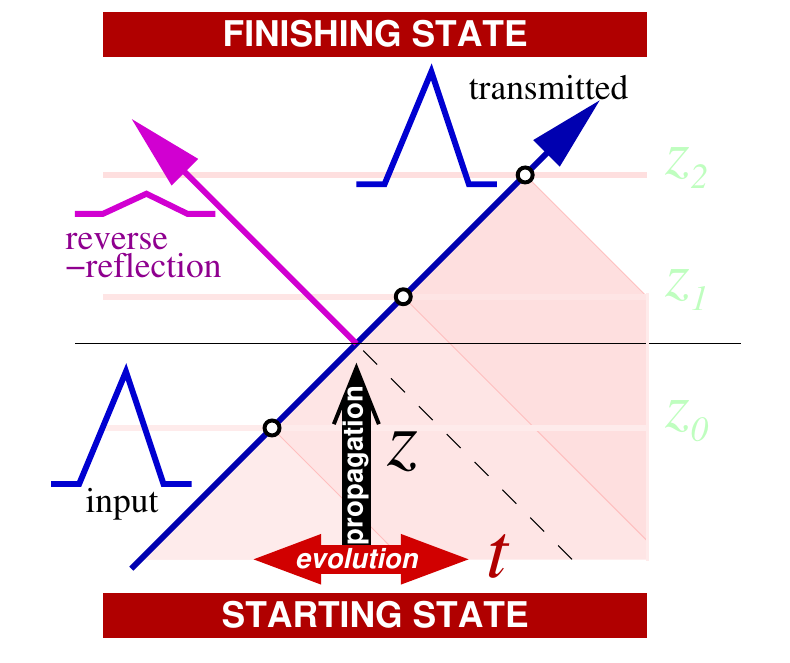}
\caption{
 Spatial propagation of waves, 
  where disturbances (or pulses)
  evolve either forward or backward in time.
 At any point in the propagation, 
  we know the full time behaviour of our wave field --
  both history and future, 
  as indicated by the pale pink horizontal lines..
 The light pink shaded triangles indicates the computational past 
  of a wave element at particular points (black circles) 
  along the path of the disturbance.
 Note that unlike the temporally propagated case
  shown in fig. \ref{fig-reflect-t}, 
  the computational past of a spatially propagated system is not the same 
  as the causal past.
 A notional interface has been added to the diagram
  to show how reflections behave --
  i.e. in an unexpected way \cite{Kinsler-2012arXiv-fbacou}.
 This is because a reflection should
  have been put in the initial conditions, 
  but was not due to an (assumed) lack of knowledge of that future behaviour.
}
\label{fig-reflect-z}
\end{figure}

The \emph{directed} position is taken when we claim
 all knowledge about how the wave the wave has moved 
 in a preferred direction along some chosen path through its environment.
This picture is shown on fig. \ref{fig-reflect-z},
 where the path is along the $z$ axis 
 from $z=-\infty$,
 through $z=0$,
 and towards $z=+\infty$.
Here we might express our knowledge of the wave-field $F$
 by writing it at the current position (e.g. $z=z_0$)
 with full time and transverse space arguments: 
 i.e. $F_{z_0}(t,x,y)$; 
 and where it is implied that we also knew $F_z(t,x,y)$ 
 at all prior positions $z<z_0$ along that path also.
This often seems a very natural thing to do, 
 especially when considering unidirectional beam propagation
 in waveguides or optical fibres.
In particular, 
 this position is made most explicit in the PSSD
 (pseudospectral spatial domain) method
 for propagating electromagnetic pulses, 
 but also in many others
 \cite{Blow-W-1989jqe,Kolesik-MM-2002prl,Genty-KKD-2007oe,Kinsler-RN-2005pra,Kinsler-2010pra-fchhg}.
Here, 
 the starting state at $z=z_i$ for some spatially propagated simulation --
 the ``computational initial conditions'' --
 could be written $F_{z_i}(t,x,y)$, 
 and
 are emphatically not the same
 as the traditional physical initial conditions at a time $t=t_i$
 (written as e.g. $F_{t_i}(x,y,z)$).
Contrast, 
 for example,
 the starting states 
 on figs. \ref{fig-reflect-t} and \ref{fig-reflect-z}.

Given some wave field state $F_z(t,x,y)$
 known at some specified path position $z_0$,
 we are at liberty to Fourier transform in $x$ and $y$ space, 
 and over time $t$, 
 but not along the propgation axis $z$.
The result will be a mixed spectrum ${F_z}(\omega,k_x,k_y)$
 as calculable for the current position $z_0$, 
 as well as prior locations $z<z_0$.
The spatial transform of this will give us 
 the transverse wavevector spectrum of the wave field, 
 which will contain both positive and negative wavevector components
 along each transverse coordinate axes.
The temporal transform, 
 using the known full past-and-future history of that transverse behaviour, 
 will contain both positive and negative frequency components.
This outcome is particularly convenient
 because (assumed) knowledge of the full time behaviour
 enables a frequency spectrum to be calculated, 
 and arbitrary temporal material response (dispersion)
 to be implemented in a numerically efficient way
 using 
 (see commentary in e.g. 
 \cite{Tyrrell-KN-2005jmo,Kinsler-2010pra-fchhg,Kinsler-2012arXiv-fbacou}
 and references therein).


For simplicity, 
 and in analogy to the comparable discussion for the traditional picture,
 imagine we are following some trajectory through space and time, 
 while counting interesting events that occur locally along that path.
In this picture,
 the spatial coordinate along our propagation axis (e.g. $z$) 
 is always increasing
 but our trajectory can move forward or back
 along the other spatial axes ($x, ~y$), 
 and even forward of back in time; 
 although it is easier to think of the case where our $x,~y$ position is fixed,
 as is time $t$.
Naturally, 
 our count of interesting events will only ever increase, 
 but rather than measuring second between events (or events per second),
 it would be in meters between events (or events per meter) --
 not a temporal period (or frequency), 
 but a \emph{spatial} interval $\lambda$
 or spatial recurrence rate (wavevector) $k_z = 2\pi/\lambda$.
Consquently, 
 any $\lambda$ (or $k_z$) estimation we might make would be a positive number. 
However, 
 the events themselves may have different characteristics:
 notably, 
 we may see objects that pass us travelling along a transverse spatial axis 
 in either direction.
Further, 
 since $F_z(t,x,y)$ contains a full time history, 
 this picture can even encode objects travelling forward or backward in time!
The differing character of events, 
 accessible to us through the spatial  and temporal knowledge in $F_z(t,x,y)$,
 allows us to impose a sign (or signs)
 when adjusting our counting total -- 
 perhaps +1 for future-going objects, 
 and -1 for past-going ones.

Having discussed how a counting/distance argument
 that gives us a strictly positive wavevector estimate
 can be converted into a signed count using temporal information, 
 let us revist the spatially propagated frequency spectrum ${F_{z_0}}(\omega)$
 of our notional wave field.
As already noted, 
 the frequency spectrum of the wave has both
 positive and negative frequency components.
Again using the well known velocity relation for waves $v=\omega/k$, 
 we can now claim that the positive frequency part
 of the spectrum represents forward (in time)
 evolving waves (with $v > 0$), 
 while the negative wavevector part represents backward (in time) 
 evolving waves (with $v < 0$).
However, 
 we have to be careful:
 even a static wave will have positive 
 and negative frequency components, 
 since for real-valued $F(t)$, 
 $F(-\omega)=F^\ast(\omega)$.
It is the changing complex phase of $F(\omega)$ that represents 
 the shift in $F(t)$ profile from one time to another.

Note that I have so far only discussed \emph{estimates}
 of the propagation axis wavevector $k_z$.
In this picture, 
 any system state exists only at a specific location (e.g. $z=z_0$)
 along its path, 
 and so of itself that state provides no information 
 about a $k_z$ at all.
Further, 
 even given our knowledge of previous states on the path,
 we cannot calculate a true $k_z$, 
 because we only know about where we have been ($z \le z_0$), 
 not where we are yet to go ($z > z_0$).
The best we might do is apply a Laplace transform over the data from 
 behind us along our path, 
 converting $z$ into a Laplace-spectral $q_z$, 
 and get hybrid spectrum ${F'_{z_0}}(\omega,k_x,k_y,q_z)$.
Nevertheless, 
 it is usually extremely advantageous to characterise the dominant $k_z$-like 
 properties of the propagation as a function of the known frequency
 and so be able to use a reference $k_z(\omega)$
 \cite{Kinsler-2012arXiv-fbacou,Kinsler-2010pra-fchhg}
 as a basis on which to simplify the propagation -- 
 perhaps by assuming it to be unidirectional.


So in the \emph{directed} picture, 
 we can directly obtain a true frequency spectrum at any point 
 in our propagation, 
 and that spectrum will have negative components.
These negative frequencies have a precise and well-defined meaning, 
 but that meaning results from the convenient
 (but physically approximate)
 decision to treat propagation as if it were along a spatial path,
 rather than forward in time.


%
\section{Causal: the past light cone}\label{S-causal}

\begin{figure}
\includegraphics[angle=-0,width=0.82\columnwidth]{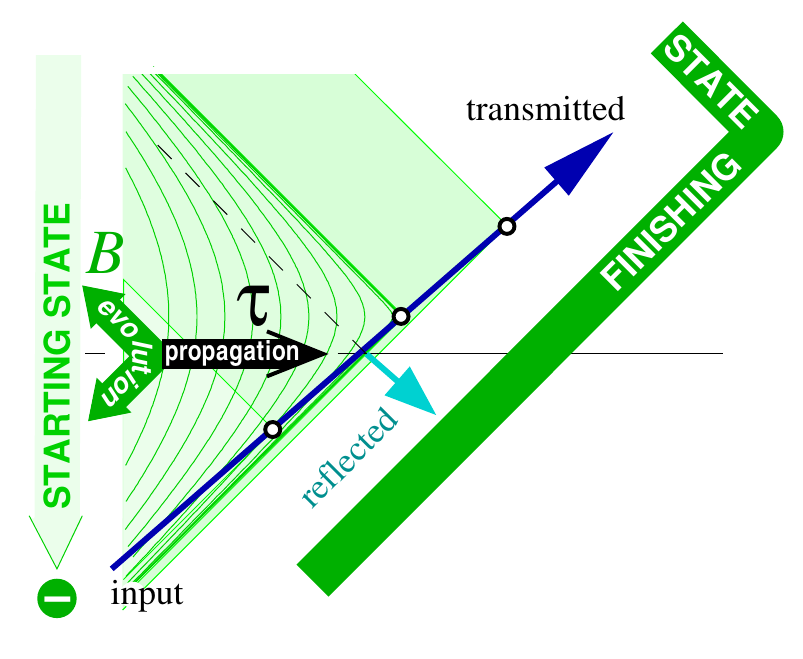}
\caption{
 Causal propagation, 
  where our knowledge only advances with the edge of
  a single point's (observer's) lightcone; 
  here I show the point travelling at slightly less
  than the maximum speed (i.e. of that of light)
  for clarity.
 At any point in the propagation, 
  we know only the behaviour in our past lightcone; 
  we show three such past lightcones, 
  one being inscribed with curves showing how the rapidity $B$ varies at
  a selection of fixed proper time $\tau$ intervals.
The initial conditions are a single point, 
  and the final state the lightcone border.
 A notional interface has been added to the diagram
  to show how reflections behave.
}
\label{fig-reflect-c}
\end{figure}

The \emph{causal} position taken when we claim
 all knowledge \emph{allowed by causal signalling} 
 about how the wave has moved 
 through its environment.
Here we might express our knowledge of some wave-field $F$
 by writing it $F(\tau,\Vec{B})$;
 where $\tau$ and $\Vec{B}$ are the proper time interval into the past
 and $\Vec{B}$ the vector ``rapidity'' needed for a signal to travel from 
 the past to our current location.
These $\tau$ and $\Vec{B}$ are constructed in a similar way 
 to Rindler coordinates, 
 and quite naturally respect the light cone.
This has been discussed in more detail elsewhere \cite{Kinsler-2011ejp}.

Since here our frequency-like quantity relates to the proper time $\tau$
 and not $t$, 
 and because our wavevector-like quantity relates to $\Vec{B}$ 
 and not $\Vec{r}$, 
 I leave any more systematic analysis to later work.
However, 
 note that $\tau$ is like time $t$ in the sense that it is one sided -- 
 we only know the past; 
 and that $\Vec{B}$ is like $\Vec{r}$ in that spans all (allowed)
 points in space.

On fig. \ref{fig-reflect-c} 
 I indicate in diagram form
 how a strictly causal simulation might proceed
 if we consider the knowledge of (an observer) at a single point
 e.g. one co-moving with a point on the wave profile.
The starting state of this causal propagation
 matches what would we would know the instant we switched on some sensors --
 i.e. only that of the observer's current location.
Here, 
 the starting state for some causally propagated simulation --
 the ``computational initial conditions'' --
 are a limited (single point) subset
 of the traditional physical initial conditions, 
 which tend to assume a much greater knowledge of the environment.
If desired, 
 these causal initial conditions could be expanded to cover
 the past light cone of the initial point.
This initial knowledge would then expand, 
 because as we propagate, 
 we would integrate our model of the system to add a ``new layer''
 to the outside of our prior past light cone 
 to get our new (updated) past light cone.

It is debatable whether this rather purist causal picture
 is of much use in a scientific setting,
 except perhaps as part of a thought experiment.
Nevertheless,
 it could be invaluable when considering
 cause and effect or the dynamical responses in specific situations, 
 such as a metamaterial element being driven in the ultrafast
 and nanoscopic regime.
On a more colourful level, 
 for those engaged in spaceship combat at relativistic velocities --
 or more likely, those science fiction authors writing about such things -- 
 it is the only view of the known environment that makes sense.

%
\section{Summary}\label{S-summary}

The above discussion shows that negative frequencies can have
 a well grounded and physical basis --  
 as long as we either are sufficiently omniscient
 and have a complete knowledge
 of the behaviour we consider relevant, 
 \emph{or} if we start from the premise that choosing spatial propagation
 is reasonable.
In such cases 
 we cannot object to the appearance of negative frequency components, 
 but it is worth noting that that we are not omniscient, 
 and that spatial propagation -- however useful -- 
 is (in practice) an approximation of reality.
This means that the concept of negative frequencies 
 must be treated with some caution 
 in any kind of dynamical situation.

%

\begin{acknowledgments}
  I would like to acknowledge the role of the recent 
 ``The Nonlinear Meeting 2014'' in Edinburgh (May 2014), 
 funded by the Max Planck Society, 
 in helping crystallize some of my thoughts on this topic, 
 and motivating me to write them up.
 I also acknowledge financial support from EPSRC \cite{EPSRC-EPK0033051}.
\end{acknowledgments}

%
\bibliography{/home/physics/_work/bibtex.bib}

\begin{thebibliography}{16}
\expandafter\ifx\csname natexlab\endcsname\relax\def\natexlab#1{#1}\fi
\expandafter\ifx\csname bibnamefont\endcsname\relax
  \def\bibnamefont#1{#1}\fi
\expandafter\ifx\csname bibfnamefont\endcsname\relax
  \def\bibfnamefont#1{#1}\fi
\expandafter\ifx\csname citenamefont\endcsname\relax
  \def\citenamefont#1{#1}\fi
\expandafter\ifx\csname url\endcsname\relax
  \def\url#1{\texttt{#1}}\fi
\expandafter\ifx\csname urlprefix\endcsname\relax\def\urlprefix{URL }\fi
\providecommand{\bibinfo}[2]{#2}
\providecommand{\eprint}[2][]{\url{#2}}

\bibitem[{\citenamefont{Rubino et~al.}(2012)\citenamefont{Rubino, McLenaghan,
  Kehr, Belgiorno, Townsend, Rohr, Kuklewicz, Leonhardt, K{\"o}nig, and
  Faccio}}]{Rubino-MKBTRKLKF-2012prl}
\bibinfo{author}{\bibfnamefont{E.}~\bibnamefont{Rubino}},
  \bibinfo{author}{\bibfnamefont{J.}~\bibnamefont{McLenaghan}},
  \bibinfo{author}{\bibfnamefont{S.~C.} \bibnamefont{Kehr}},
  \bibinfo{author}{\bibfnamefont{F.}~\bibnamefont{Belgiorno}},
  \bibinfo{author}{\bibfnamefont{D.}~\bibnamefont{Townsend}},
  \bibinfo{author}{\bibfnamefont{S.}~\bibnamefont{Rohr}},
  \bibinfo{author}{\bibfnamefont{C.~E.} \bibnamefont{Kuklewicz}},
  \bibinfo{author}{\bibfnamefont{U.}~\bibnamefont{Leonhardt}},
  \bibinfo{author}{\bibfnamefont{F.}~\bibnamefont{K{\"o}nig}},
  \bibnamefont{and} \bibinfo{author}{\bibfnamefont{D.}~\bibnamefont{Faccio}},\\
  \bibinfo{journal}{Phys. Rev. Lett.} \textbf{\bibinfo{volume}{108}},
  \bibinfo{pages}{253901} (\bibinfo{year}{2012}),\\
  \XDOI{10.1103/PhysRevLett.108.253901}.

\bibitem[{\citenamefont{Conforti et~al.}(2013)\citenamefont{Conforti,
  Westerberg, Baronio, Trillo, and Faccio}}]{Conforti-WBTF-2013pra}
\bibinfo{author}{\bibfnamefont{M.}~\bibnamefont{Conforti}},
  \bibinfo{author}{\bibfnamefont{N.}~\bibnamefont{Westerberg}},
  \bibinfo{author}{\bibfnamefont{F.}~\bibnamefont{Baronio}},
  \bibinfo{author}{\bibfnamefont{S.}~\bibnamefont{Trillo}}, 
  \bibinfo{author}{\bibfnamefont{D.}~\bibnamefont{Faccio}},\\
  \bibinfo{journal}{Phys. Rev. A} \textbf{\bibinfo{volume}{88}},
  \bibinfo{pages}{013829} (\bibinfo{year}{2013}),\\
  \XDOI{10.1103/PhysRevA.88.013829}.

\bibitem[{\citenamefont{Rousseaux et~al.}(2008)\citenamefont{Rousseaux, Mathis,
  Ma{\"{\i}}ssa, Philbin, and Leonhardt}}]{Rousseaux-MMPL-2008njp}
\bibinfo{author}{\bibfnamefont{G.}~\bibnamefont{Rousseaux}},
  \bibinfo{author}{\bibfnamefont{C.}~\bibnamefont{Mathis}},
  \bibinfo{author}{\bibfnamefont{P.}~\bibnamefont{Ma{\"{\i}}ssa}},
  \bibinfo{author}{\bibfnamefont{T.~G.} \bibnamefont{Philbin}},
  \bibinfo{author}{\bibfnamefont{U.}~\bibnamefont{Leonhardt}},\\
  \bibinfo{journal}{New J. Phys.} \textbf{\bibinfo{volume}{10}},
  \bibinfo{pages}{053015} (\bibinfo{year}{2008}),\\
  \XDOI{10.1088/1367-2630/10/5/053015}.

\bibitem[{\citenamefont{Kinsler}(2012{\natexlab{a}})}]{Kinsler-2012arXiv-fbaco%
u}
\bibinfo{author}{\bibfnamefont{P.}~\bibnamefont{Kinsler}}, \\
  \bibinfo{journal}{J. Phys. Commun.} \textbf{\bibinfo{volume}{10}},
  \bibinfo{pages}{025011} (\bibinfo{year}{2018}),\\
  \XDOI{10.1088/2399-6528/aaa85c},\\
  \XARXIV{1202.0714}.

\bibitem[{\citenamefont{Kinsler}(2010)}]{Kinsler-2010pra-fchhg}
\bibinfo{author}{\bibfnamefont{P.}~\bibnamefont{Kinsler}},\\
  \bibinfo{journal}{Phys. Rev. A} \textbf{\bibinfo{volume}{81}},
  \bibinfo{pages}{013819} (\bibinfo{year}{2010}),\\
  \XDOI{10.1103/PhysRevA.81.013819},\\
  \XARXIV{0810.5689}.


\bibitem[{\citenamefont{Kinsler}(2012{\natexlab{b}})}]{Kinsler-2012arxiv-fbrad}
\bibinfo{author}{\bibfnamefont{P.}~\bibnamefont{Kinsler}}, \\
  \bibinfo{title}{Directional pulse propagation in beam, rod, pipe, and disk geometries}\\
  (\bibinfo{year}{2012}),\\
  \XARXIV{1210.6794}.

\bibitem[{\citenamefont{Yee}(1966)}]{Yee-1966tap}
\bibinfo{author}{\bibfnamefont{K.~S.} \bibnamefont{Yee}},\\
  \bibinfo{journal}{IEEE Trans. Antennas Propagat.}
  \textbf{\bibinfo{volume}{14}}, \bibinfo{pages}{302} (\bibinfo{year}{1966}),\\
  \XDOI{10.1109/TAP.1966.1138693}

\bibitem[{\citenamefont{Oskooi et~al.}(2010)\citenamefont{Oskooi, Roundy,
  Ibanescu, Bermel, Joannopoulos, and Johnson}}]{Oskooi-RIBJJ-2010cpc}
\bibinfo{author}{\bibfnamefont{A.~F.} \bibnamefont{Oskooi}},
  \bibinfo{author}{\bibfnamefont{D.}~\bibnamefont{Roundy}},
  \bibinfo{author}{\bibfnamefont{M.}~\bibnamefont{Ibanescu}},
  \bibinfo{author}{\bibfnamefont{P.}~\bibnamefont{Bermel}},
  \bibinfo{author}{\bibfnamefont{J.~D.} \bibnamefont{Joannopoulos}},
  \bibnamefont{and} \bibinfo{author}{\bibfnamefont{S.~G.}
  \bibnamefont{Johnson}}, \\
  \bibinfo{journal}{Comput. Phys. Commun.}
  \textbf{\bibinfo{volume}{181}}, \bibinfo{pages}{687} (\bibinfo{year}{2010}),\\
  \XDOI{10.1016/j.cpc.2009.11.008}.

\bibitem[{\citenamefont{Kinsler}(2012{\natexlab{a}})}]{Kinsler-2018jo-d2owe}
\bibinfo{author}{\bibfnamefont{P.}~\bibnamefont{Kinsler}}, \\
  \bibinfo{journal}{J. Opt.} \textbf{\bibinfo{volume}{20}},
  \bibinfo{pages}{025502} (\bibinfo{year}{2018}),\\
  \XDOI{10.1088/2040-8986/aaa0fc},\\
  \XARXIV{1501.05569}.

\bibitem[{\citenamefont{Blow and Wood}(1989)}]{Blow-W-1989jqe}
\bibinfo{author}{\bibfnamefont{K.~J.} \bibnamefont{Blow}} \bibnamefont{and}
  \bibinfo{author}{\bibfnamefont{D.}~\bibnamefont{Wood}},\\
  \bibinfo{journal}{IEEE J. Quantum Electronics} \textbf{\bibinfo{volume}{25}},
  \bibinfo{pages}{2665} (\bibinfo{year}{1989}),\\
  \XDOI{10.1109/3.40655}.

\bibitem[{\citenamefont{Kolesik et~al.}(2002)\citenamefont{Kolesik, Moloney,
  and Mlejnek}}]{Kolesik-MM-2002prl}
\bibinfo{author}{\bibfnamefont{M.}~\bibnamefont{Kolesik}},
  \bibinfo{author}{\bibfnamefont{J.~V.} \bibnamefont{Moloney}},
  \bibnamefont{and} \bibinfo{author}{\bibfnamefont{M.}~\bibnamefont{Mlejnek}},\\
  \bibinfo{journal}{Phys. Rev. Lett.} \textbf{\bibinfo{volume}{89}},
  \bibinfo{pages}{283902} (\bibinfo{year}{2002}),\\
  \XDOI{10.1103/PhysRevLett.89.283902}.

\bibitem[{\citenamefont{Genty et~al.}(2007)\citenamefont{Genty, Kinsler,
  Kibler, and Dudley}}]{Genty-KKD-2007oe}
\bibinfo{author}{\bibfnamefont{G.}~\bibnamefont{Genty}},
  \bibinfo{author}{\bibfnamefont{P.}~\bibnamefont{Kinsler}},
  \bibinfo{author}{\bibfnamefont{B.}~\bibnamefont{Kibler}}, \bibnamefont{and}
  \bibinfo{author}{\bibfnamefont{J.~M.} \bibnamefont{Dudley}}, \\
  \bibinfo{journal}{Opt. Express} \textbf{\bibinfo{volume}{15}},
  \bibinfo{pages}{5382} (\bibinfo{year}{2007}), \\
  \XDOI{10.1364/OE.15.005382}.

\bibitem[{\citenamefont{Kinsler et~al.}(2005)\citenamefont{Kinsler, Radnor, and
  New}}]{Kinsler-RN-2005pra}
\bibinfo{author}{\bibfnamefont{P.}~\bibnamefont{Kinsler}},
  \bibinfo{author}{\bibfnamefont{S.~B.~P.} \bibnamefont{Radnor}},
  \bibnamefont{and} \bibinfo{author}{\bibfnamefont{G.~H.~C.}
  \bibnamefont{New}}, \\
  \bibinfo{journal}{Phys. Rev. A}
  \textbf{\bibinfo{volume}{72}}, \bibinfo{pages}{063807}
  (\bibinfo{year}{2005}), \\
  \XDOI{10.1103/PhysRevE.75.066603},\\
  \XARXIV{physics/0611215}.

\bibitem[{\citenamefont{Tyrrell et~al.}(2005)\citenamefont{Tyrrell, Kinsler,
  and New}}]{Tyrrell-KN-2005jmo}
\bibinfo{author}{\bibfnamefont{J.~C.~A.} \bibnamefont{Tyrrell}},
  \bibinfo{author}{\bibfnamefont{P.}~\bibnamefont{Kinsler}}, 
  \bibinfo{author}{\bibfnamefont{G.~H.~C.} \bibnamefont{New}}, \\
  \bibinfo{journal}{J. Mod. Opt.} \textbf{\bibinfo{volume}{52}},
  \bibinfo{pages}{973} (\bibinfo{year}{2005}), \\
  \XDOI{10.1080/09500340512331334086}.

\bibitem[{\citenamefont{Kinsler}(2011)}]{Kinsler-2011ejp}
\bibinfo{author}{\bibfnamefont{P.}~\bibnamefont{Kinsler}}, \\
  \bibinfo{journal}{Eur. J. Phys.} \textbf{\bibinfo{volume}{32}},
  \bibinfo{pages}{1687} (\bibinfo{year}{2011}), \\
  \XDOI{10.1088/0143-0807/32/6/022}, \\
  \XARXIV{1106.1792} (a longer and updated version).

\bibitem[{\citenamefont{Kinsler et~al.}(2006)\citenamefont{Kinsler, New, and
  Tyrrell}}]{Kinsler-NT-2006-phnlo}
\bibinfo{author}{\bibfnamefont{P.}~\bibnamefont{Kinsler}},
  \bibinfo{author}{\bibfnamefont{G.~H.~C.} \bibnamefont{New}}, 
  \bibnamefont{and} \bibinfo{author}{\bibfnamefont{J.~C.~A.}
  \bibnamefont{Tyrrell}}, \\
  (\bibinfo{year}{2006}), \\
  \XARXIV{physics/0611213}.

\bibitem{EPSRC-EPK0033051}
  \bibinfo{author}{\bibfnamefont{M.~W.}~\bibnamefont{McCall}},
  \bibinfo{author}{\bibfnamefont{P.}~\bibnamefont{Kinsler}},\\
   EPSRC grant number EP/K003305/1.

\bibitem[{\citenamefont{Gulley and Dennis}(2010)}]{Gulley-D-2010pra}
\bibinfo{author}{\bibfnamefont{J.~R.} \bibnamefont{Gulley}} \bibnamefont{and}
  \bibinfo{author}{\bibfnamefont{W.~M.} \bibnamefont{Dennis}},\\
  \bibinfo{journal}{Phys. Rev. A} \textbf{\bibinfo{volume}{81}},
  \bibinfo{pages}{033818} (\bibinfo{year}{2010}),\\
  \XDOI{10.1103/PhysRevA.81.033818}.

\bibitem[{\citenamefont{Kinsler and New}(2005)}]{Kinsler-N-2005pra}
\bibinfo{author}{\bibfnamefont{P.}~\bibnamefont{Kinsler}} \bibnamefont{and}
  \bibinfo{author}{\bibfnamefont{G.~H.~C.} \bibnamefont{New}},\\
  \bibinfo{journal}{Phys. Rev. A} \textbf{\bibinfo{volume}{72}},
  \bibinfo{pages}{033804} (\bibinfo{year}{2005}), \\
  \XDOI{10.1103/PhysRevA.72.033804},\\
  \XARXIV{physics/0606111}.

\bibitem[{\citenamefont{Stolen et~al.}(1989)\citenamefont{Stolen, Gordon,
  Tomlinson, and Haus}}]{Stolen-GTH-1989josab}
\bibinfo{author}{\bibfnamefont{R.~H.} \bibnamefont{Stolen}},
  \bibinfo{author}{\bibfnamefont{J.~P.} \bibnamefont{Gordon}},
  \bibinfo{author}{\bibfnamefont{W.~J.} \bibnamefont{Tomlinson}},
  \bibnamefont{and} \bibinfo{author}{\bibfnamefont{H.~A.} \bibnamefont{Haus}},\\
  \bibinfo{journal}{J. Opt. Soc. Am. B} \textbf{\bibinfo{volume}{6}},
  \bibinfo{pages}{1159} (\bibinfo{year}{1989}),\\
  \XDOI{10.1364/JOSAB.6.001159}.

\bibitem[{\citenamefont{Kinsler}(2010{\natexlab{b}})}]{Kinsler-2010pra-lfiadc}
\bibinfo{author}{\bibfnamefont{P.}~\bibnamefont{Kinsler}},\\
  \bibinfo{journal}{Phys. Rev. A} \textbf{\bibinfo{volume}{82}},
  \bibinfo{pages}{055804} (\bibinfo{year}{2010}{\natexlab{b}}),\\
  \XDOI{10.1103/PhysRevA.81.013819},\\
  \XARXIV{1008.2088}.

\bibitem[{\citenamefont{Kinsler}(2019{\natexlab{a}})}]{Kinsler-2019arxiv-spatype}
\bibinfo{author}{\bibfnamefont{P.}~\bibnamefont{Kinsler}}, \\
  \bibinfo{title}{An introduction to spatial dispersion}\\
  (\bibinfo{year}{2019}),\\
  \XARXIV{1904.11957}.

\end{thebibliography}

%
\section*{Appendix: Time response models}\label{S-timer}

The time response of a propagation medium is handled rather 
 differently in the temporal and spatial propagation approaches.
Consider a set of material properties $n_i$
 that follow some kind of dynamical (temporal) response models
 defined by a differential equation.
The differential equations for each $n_i$ will
 depends on both the local material states and that of the local field:
~
\begin{align}
  \partial_t^k n_i &= r_i(\{n_j\},F)
.
\label{eqn-S-timer-response}
\end{align}
Typical examples might be
 nonlinear response delays \cite{Blow-W-1989jqe},
 a free carrier density model \cite{Gulley-D-2010pra}, 
 a Raman response \cite{Stolen-GTH-1989josab,Kinsler-N-2005pra}, 
 or even just a Drude or Lorentz oscillator \cite{Kinsler-2011ejp}
 giving rise to dispersion.
The requirement here for the model to be properly causal 
 is that the order of the time derivative $k$
 is greater than any other time derivative parts present in 
 the response function (or operator) $r_i$ \cite{Kinsler-2010pra-lfiadc}.

\noindent
$\blacktriangleright$ ~
In the traditional time propagated picture, 
 each spatial location needs to not only know its field state $F(\Vec{r})$, 
 but also its material property state(s) $n_i(\Vec{r})$.
In some situations, 
 values of $F$ or $n_i$ from earlier times --
 i.e. values from the computational past of the simulation --
 may need to be stored for use
 in subsequent computations.
Further, 
 as the propagation proceeds step-by-step in time,
 \emph{both} the field $F(\Vec{r})$
 and current material properties $n_i(\Vec{r})$ 
 need to be updated.

\noindent
$\blacktriangleright$ ~
In the (directed) space propagated picture, 
 each currently held state of the field holds the time history 
 of each point.
This means that it is never necessary to store the computational past
 of the simulation,
 since the values for earlier times are already incorporated into
 the current computational state $F(t,x,y)$.
We can simply solve 
 equations such as eqn. \eqref{eqn-S-timer-response}
 by directly integrating them from the distant past 
 up to the desired time,
 using only that current state information.
And, 
 as already mentioned, 
 if the response model is linear and has a known frequency response, 
 as in the case of typical dispersive behaviours, 
 it can be applied quickly and efficiently as a simple pahse shift
 applied to the frequency spectrum --
 a process requiring only two Fourier transforms and a multiplication.

%
\section*{Appendix: Nonlinear optics and negative frequencies}\label{S-nlo}

There are some specific issues relating to nonlinear optics (NLO) and
 negative frequencies which bear additional examination.
Consider the nonlinear Schr\"odinger equation (NLSE)
 as derived from Maxwell's equations
 in the 1+1D regime of $(t,z)$,
 with some sources of dispersion and a perturbative third order (Kerr)
 nonlinearity.
Generally in NLO it is written in the convenient (i.e. directed)
 spatially propagated picture,
 in a unidirectional approximation 
 with fields evolving forward in time only,
 as the zNLSE \cite{Kinsler-2010pra-fchhg}
~
\begin{align}
  \partial_z E^+(t)
&=
  \imath K \, E^+(t)
 +
  \sum_j \beta_j \, \partial_t^j \, E^+(t)
 +
  \imath \chi_z \left| E^+(t) \right|^2 E^+(t)
,
\label{eqn-zNLS}
\end{align}
but note that it is also possible to derive a temporally propagated version, 
 for the case of a unidirectional field evolving forward in space only, 
 as the tNLSE\footnote{Author's derivation, unpublished}
~
\begin{align}
  \partial_t D_+(z)
&=
  \imath \Omega \, D_+(z)
 +
  \sum_j \gamma_j \, \partial_z^j \, D_+(z)
 +
  \imath \chi_t \left| D_+(z) \right|^2 D_+(z)
.
\label{eqn-tNLS}
\end{align}

Note that in the ``directed'' spatial zNLSE eqn. \eqref{eqn-zNLS}
 all sources of dispersion --
 whether due to time-dependent material response or the geometric properties
 of a confining waveguide -- 
 are treated as if they were \emph{solely} due 
 to a time-dependent response
 (by means of the power series in time derivatives).
In contrast, 
 in the ``traditional'' temporal zNLSE eqn. \eqref{eqn-tNLS}, 
 all sources of dispersion in the tNLSE
 are treated as if they were \emph{solely} due 
 to the local geometric properties of that material
 i.e. by spatial dispersion \cite{Kinsler-2019arxiv-spatype}; 
 typically this is expressed as 
 a power series in either spatial derivatives or wavevector.
However, 
 given that most dispersions are rather weak in applications NLSE propagation, 
 either picture can plausibly be assumed to be able to
 treat dispersion accurately enough for such purposes
 (although see \cite{Kinsler-2018jo-d2owe}).

In both these cases we can see that it would be useful to Fourier transform
 the field to get temporal and spatial spectra 
 $\tilde{E}^+(\omega)$ and $\bar{D}_+(k)$ respectively, 
 because then the dispersion model can be directly applied
 as a simple polynomial phase shift, 
 i.e. either of 
~
\begin{align}
  \beta(\omega) &= \sum_j \beta_j  \,.\, \left( -\imath \omega \right)^j
\\
  \gamma(k)     &= \sum_j \gamma_j \,.\, \left(  \imath k \right)^j
\end{align}

It is often stated that a Kerr nonlinearity
 causes third harmonic generation, 
 but we now also need to consider what this generation
 is a ``third harmonic'' of.
In the usual zNLSE model of eqn. \eqref{eqn-zNLS},
 an existing harmonic field 
 $E^+(t) \simeq E^+_0 e^{-\imath \omega_1 t} + \textrm{c.c.}$
 leads to not only some self phase modulation (SPM)
 but also third harmonic frequency generation at $\omega_3 = 3 \omega_1$.
In the tNLSE model of eqn. \eqref{eqn-tNLS},
 an existing harmonic field 
 $D_+(z) \simeq D_+^0 e^{\imath k_1 z} + \textrm{c.c.}$
 leads to not only some self phase modulation (SPM)
 but also third harmonic wavevector generation at $k_3 = 3 k_1$.

For clarity,
 let us write this out as carefully as possible, 
 in the case where we spilt the real-valued fields $E^+$ or $D_+$
 into complex conjugate halves, 
 in the style of \cite{Kinsler-NT-2006-phnlo}.
With 
 $E^+(t) = E'(t) e^{-\imath \omega_1 t} + E'^\ast(t) e^{+\imath \omega_1 t}$, 
 the zNLSE equation can be partitioned into two complex conjugate halves.
We can also dispense with the 
 exponential oscillation by setting $\omega_1=0$ --  
 although it is often useful to include such carrier oscillations
 when the fields are narrow band,
 having them written explicitly rather than implicitly
 is not required.
The two complex conjugate zNLSE equations are
~
\begin{align}
  \partial_z E'(t) e^{-\imath \omega_1 t}
&=
  \imath K \, E'(t) e^{-\imath \omega_1 t}
 +
  \sum_j \beta_j \, \partial_t^j \, E'(t) e^{-\imath \omega_1 t}
\nonumber
\\
&\quad
 +
  \imath \chi_z 
  \left[
    E'(t)^3 e^{-3 \imath \omega_1 t}
   +
    3 E'(t)^2 \left[ E'(t) \right]^\ast e^{-\imath \omega_1 t}
  \right]
,
\label{eqn-zNLS-cc1}
\\
  \partial_z E'^\ast(t) e^{+\imath \omega_1 t}
&=
  \imath K \, E'^\ast(t) e^{+\imath \omega_1 t}
 +
  \sum_j \beta_j \, \partial_t^j \, E'^\ast(t) e^{+\imath \omega_1 t}
\nonumber
\\
&\quad
 +
  \imath \chi_z 
  \left[
    E'^\ast(t)^3 e^{+3 \imath \omega_1 t}
   +
    3 E'^\ast(t)^2 \left[ E'^\ast(t) \right]^\ast e^{+\imath \omega_1 t}
  \right]
.
\label{eqn-zNLS-cc2}
\end{align}
The sum of these two equations is just eqn. \eqref{eqn-zNLS}.
Further, 
 since they are exact complex conjugates of one another, 
 we can solve just one version, 
 which automatically gives us a solution to the other, 
 and hence the solution to the real valued field $E^+(t)$.

Note in particular that the nonlinearity
 in this exact mathematical re-expression 
 of the zNLSE equation
 drives only resonant or third-harmonic frequencies.
There is no explicit coupling to negative frequencies, 
 which can be seen most clearly when assuming 
 finite $\omega_1$ and a constant $E'$; 
 then, 
 eqn. \eqref{eqn-zNLS-cc1} does not drive negative frequencies of itself.
Nevertheless, 
 although the complex $E'(\omega)$ might start with content
 solely in positive frequencies, 
 its frequency bandwidth is not restricted.
Consequently
 after propagating some distance it may well have evolved
 into a field whose spectrum is very wide, 
 perhaps even extending past the origin.
In this state, 
 multi-frequency interactions that drive the field at negative frequencies
 could indeed occur as a result of the nonlinearity --
 remember that if represented in the frequency domain, 
 the cubic nonlinear term transforms into a double convolution
 over the entire frequency spectra.

However, 
 we could,
 if we wanted, 
 make such ``negative frequency'' driving terms
 appear explicitly by repartitioning the whole nonlinear term
 into different complex conjugate halves,
 e.g. by writing not eqn. \eqref{eqn-zNLS-cc1} but
 ~
\begin{align}
  \partial_z E'(t) e^{-\imath \omega_1 t}
&=
  \imath K \, E'(t) e^{-\imath \omega_1 t}
 +
  \sum_j \beta_j \, \partial_t^j \, E'(t) e^{-\imath \omega_1 t}
\nonumber
\\
&\quad
 +
  \imath \chi_z 
  \left[
    E'(t)^3 e^{-3 \imath \omega_1 t}
   +
    \frac{3}{2} E'(t)^2 \left[ E'(t) \right]^\ast   e^{-\imath \omega_1 t}
  \right.
\nonumber
\\
&\qquad\qquad\qquad
  \left.
   +
    \frac{3}{2} \left[ E'(t) \right]^{\ast 2} E'(t) e^{+\imath \omega_1 t}
  \right]
,
\label{eqn-zNLS-cc1-v2}
\end{align}
 and there is of course also a matching complex congugate counterpart
 of eqn. \eqref{eqn-zNLS-cc2}, 
 and the sum of both will be equal to the original zNLSE equation.
Whether or not this representation containing 
 explicit driving of negative frequencies
 might be useful is another matter, 
 but it is certainly possible to (re)express the mathematical model 
 in order to construct them.
But, 
 apart from specific numerical difficulties that might occur when solving
 these propagation equations, 
 the solutions gained from either form should be identical:
 \emph{both are just different ways of representing the same physical model}.

Naturally one can apply the same method to the tNLSE as well, 
 setting 
 $D_+(z) = D'(z) e^{\imath k_1 t} + D'^\ast(z) e^{-\imath k_1 z}$, 
 and arriving at 
 two complex conjugate equations
~
\begin{align}
  \partial_z D'(z) e^{+\imath k_1 z}
&=
  \imath K \, D'(z) e^{+\imath k_1 z}
 +
  \sum_j \gamma_j \, \partial_z^j \, D'(z) e^{+\imath k_1 z}
\nonumber
\\
&\quad
 +
  \imath \chi_t
  \left[
    D'(z)^3 e^{+3 \imath k_1 z}
   +
    3 D'(z)^2 \left[ D'(z) \right]^\ast e^{+\imath k_1 z}
  \right]
,
\label{eqn-tNLS-cc1}
\\
  \partial_z D'^\ast(z) e^{-\imath k_1 z}
&=
  \imath K \, D'^\ast(z) e^{-\imath k_1 z}
 +
  \sum_j \gamma_j \, \partial_z^j \, D'^\ast(z) e^{-\imath k_1 z}
\nonumber
\\
&\quad
 +
  \imath \chi_t 
  \left[
    D'^\ast(z)^3 e^{-3 \imath k_1 z}
   +
    3 D'^\ast(z)^2 \left[ D'^\ast(z) \right]^\ast e^{-\imath k_1 z}
  \right]
.
\end{align}
As we should expect, 
 these two equations sum to eqn. \eqref{eqn-tNLS}.
Also,
 here there again is no explicit coupling to opposite parts
 of the wavevector spectra $D'(k)$
 in each equation, 
 although (as before) we might repartition the nonlinear term
 to construct it, 
 if we so desired.

\end{document}